\documentclass{article}
\usepackage{amsmath}
\usepackage{url}

\usepackage[margin=1in]{geometry}
\usepackage{graphicx}
\usepackage{amsmath}
\usepackage{epic}
\usepackage{latexsym}
\usepackage{epsfig}
\usepackage{amssymb}
\usepackage{enumerate}
\usepackage[mathscr]{euscript}
\usepackage{color}
\usepackage{indentfirst}
\usepackage{mathtools}
\newtheorem{theorem}{Theorem}[section]
\newtheorem{lemma}[theorem]{Lemma} 

\newtheorem{proposition}[theorem]{Proposition}

\newcommand{\PP}{{\mathbb P}}
\newcommand{\EE}{{\mathbb E}}

\newcommand{\cC}{{\mathcal C}}
\newcommand{\cG}{{\mathcal G}}

\begin{document}

\title{Tracing evolutionary links between species}
\author{Mike Steel\\
Biomathematics Research Centre, \\
University of Canterbury, Christchurch, NZ}

\bigskip

\maketitle

\begin{abstract}

The idea that all  life on earth traces back to a common beginning dates back at least  to Charles Darwin's {\em Origin of Species}.
Ever since, biologists have tried to piece together parts of this `tree of life'  based on what we can observe today:  fossils, and the 
evolutionary signal that is present in the genomes and phenotypes of different organisms.  Mathematics has played a key role in helping transform genetic data into phylogenetic (evolutionary) trees and networks. Here, I will explain some of the central concepts and basic results in phylogenetics, which benefit from several branches of mathematics, including combinatorics, probability and algebra. 

\end{abstract}

\bigskip

\section{What is phylogenetics?}
All living organisms on earth harbor within their DNA a signature of their evolutionary heritage.  By studying patterns and differences between the genetic makeup of 
different species, molecular biologists are able to piece together parts of the story of how life today traces back a common origin. In this way, many basic  questions can be answered. When did animals and plants diverge? Are fungi more closely related to plants or animals? How and when did photosynthesis arise? What is the closest living animal to the whales? Does speciation occur in bursts or at a steady rate?  Other topics are proving more difficult to resolve -- for example, deciphering the earliest history of life on earth.

Similar questions arise for evolutionary processes in other fields such as epidemiology (e.g. the relationship between different strains of influenza or HIV) and linguistics (e.g. how
languages diverged from one another over time).  In all these fields, the analysis relies on an underlying mathematical theory, grounded in combinatorics, algebra, and stochastic processes, with the concept of an evolutionary tree  as a unifying object.

In this article, I describe a cross-section of some of the key concepts in `phylogenetics', which is the theory of reconstructing and analyzing trees and networks from data observed at the present.  I describe some combinatorial features of phylogenetic trees, namely their encoding by set systems,  their enumeration, their generation under random models of evolution,  and the way in which they can `perfectly' display discrete data.  I then focus on tree reconstruction from  data (discrete or distance-based), which may not perfectly fit a tree. Such imperfect data can occur when data `evolve' along the branches of the tree under a random Markov model. I end by outlining how tree reconstruction is possible from this evolved data, but the choice of method requires care, to avoid falling into a `zone' of statistical inconsistency.

\section{Hierarchies and phylogenetic trees.}
The 18th century Swedish taxonomist Carl Linneaus  noticed that much of the living world can be nicely organised into a `hierarchy' in which groups of living organisms are either disjoint or nested \cite{lin}. For example, cats and dogs comprise disjoint classes of organisms, but both are subsets of the class of mammals.  Formally,  a  {\em hierarchy} $H$ on a finite set $X$ is a collection of subsets of $X$ with the property that any two elements of $H$ are either 
nested (one is contained in the other) or disjoint. It will also be convenient here to  require that any hierarchy on $X$ contains the set $X$ and all its singleton subsets.  Thus $H$ forms a hierarchy if it satisfies the two properties:
\begin{itemize}
\item[{\bf H1:}]
 For any two sets $A,B \in H$ we have $A \cap B \in \{A, B, \emptyset\}$; and
\item[{\bf H2:}]  $H$ contains the entire set $X$, and each singleton set $\{x\}$ for all $x\in X$. 
\end{itemize}
The second condition is harmless: if $H$ is any collection of  sets that satisfies {\bf H1}, we can always add the extra elements mentioned by {\bf H2} without violating  {\bf H1}.

To connect hierarchies with trees, recall first that a {\em  tree} $T$ is a connected graph $(V,E)$ with no cycles. Often we will deal with rooted trees for which the edges are all directed away from some root vertex,  and so each vertex has an `in-degree' and `out-degree'.  We  first define a {\em rooted phylogenetic $X$-tree} to be a tree $T$ in which:
\begin{itemize}
\item $X$ is the set of leaves (vertices of out-degree 0);
\item  all the arcs (directed edges) are directed away from some root vertex $\rho$;
\item every non-leaf vertex has out-degree at least 2.

\end{itemize}
\begin{figure}[htb]
\centering
\includegraphics[scale=1.0]{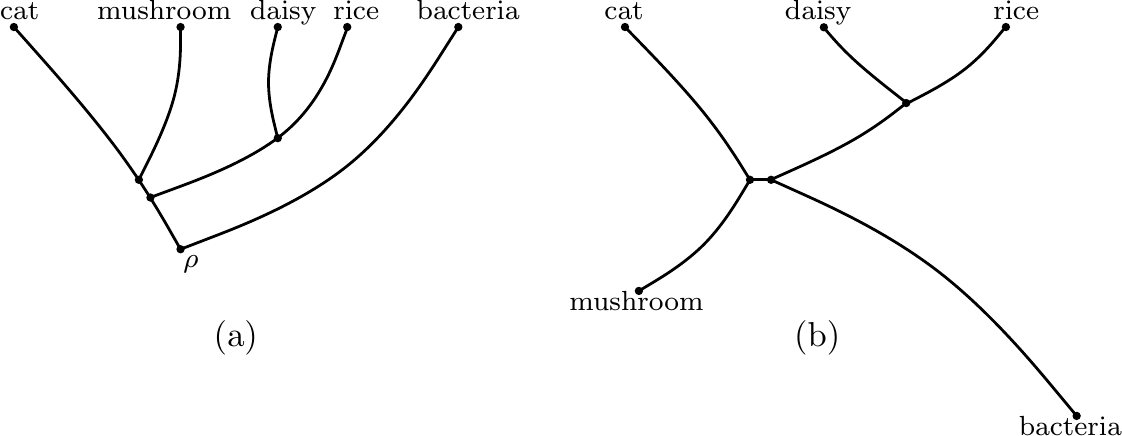}
\caption{(a) A rooted phylogenetic $X$-tree, with root $\rho$. (b) The associated unrooted tree obtained by suppressing the root vertex}
\label{tree1}
\end{figure}

Fig.~1(a) shows a simple biological example of a phylogenetic $X$-tree for a set $X$ of five species; it reveals one relationship that is perhaps surprising to most non-biologists:  genetic data indicates that fungi are more closely related genealogically to animals than to plants. The interior vertices of a phylogenetic tree represent hypothetical ancestral species, with the root $\rho$ being the `most recent common ancestor' of the species at the leaves.  

We will think of two rooted $X$-trees as equivalent if they are isomorphic as rooted trees by an isomorphism that
is the identity on $X$ (i.e. trees are equivalent up to relabelling of the non-leaf vertices). 
Given a vertex $v$ of $T$, the {\em cluster} associated with $v$ is the subset of $X$ that  becomes separated from the root upon deletion of $v$.  For  the tree
in Fig~\ref{tree1}(a), the sets $\{{\rm cat, mushroom} \}, \{{\rm daisy, rice}\}$ and $\{{\rm cat, mushroom, daisy, rice}\}$ are clusters.

Any collection $\cC$ of subsets of $X$ forms a directed graph,  sometimes called the  `cover digraph' of 
$\cC$.  The vertices of this graph are the elements of $\cC$ and we place an arc from $B \in \cC$ to $A \in \cC$ precisely if $B$ covers $A$ (i.e. $A \subset B$ and there is no set $C \in \cC$ with $A \subset C \subset B$).   Now, the clusters of any phylogenetic $X$-tree $T$ form a hierarchy on $X$, and the cover digraph of this hierarchy  is isomorphic  to $T$ under the map that sends each vertex of $T$ to the cluster associated with that vertex. Moreover, every hierarchy can be realized in this way, and nonequivalent phylogenetic trees give rise to different hiearchies. In other words, we have the following fundamental bijective correspondence between hierarchies and rooted phylogenetic $X$-trees (up to equivalence) \cite{edm}.
\begin{lemma}
\label{equiv}
\mbox{ }
A collection of subsets $H$ of $X$  is a hierarchy if and only if $H$ is the set of clusters of some rooted phylogenetic $X$-tree $T$. Moreover, $T$ is  unique up to equivalence. 
\end{lemma}

The maximal hierarchies correspond to rooted phylogenetic $X$-trees in which every non-leaf vertex of $T$ has out-degree 2, which are called  {\em binary} trees (the tree in Fig. 1(a) is an example). We will see shortly that such trees have exactly $2n-1$ vertices where $n=|X|$, and so (by Lemma~\ref{equiv}) this is the size of the largest hierarchy on a set of size $n$.
Biologists often prefer trees to be binary since they show just one lineage splitting off at a time;  by contrast a vertex of out-degree three or more represents what biologists call a `polytomy' (usually interpreted as uncertainty about the order of speciation events, rather than certainty about a  sudden speciation event into multiple lineages).

The utility of viewing a rooted phylogenetic tree as a set system (a hierarchy) is illustrated by two questions biologists often face.  Suppose we have a collection of different trees that estimate the evolutionary history of the same set of taxa. These trees might have been constructed by comparing  genetic data across these species, but different choices of which genetic data to use (e.g. different genes) could have resulted in different tree estimates.  In other words, while there might be one underlying and unknown `true' species tree that we wish to infer, the phylogenetic trees constructed from data will typically be merely imperfect estimates of this tree, since the data evolves randomly, a topic we will discuss later. So two problems arise: 
\begin{itemize}
\item
How can we compare different phylogenetic $X$-trees?  
\item 
Can we combine different phylogenetic $X$-trees into some `consensus' tree?  
\end{itemize}
The  hierarchy link provides a very simple solution to both questions.   First, observe that we can define a distance $d$ between any two rooted phylogenetic $X$-trees $T$ and $T'$  by taking $d(T,T')$ to be the number of clusters which are present in one but not both of the trees $T$ and $T'$. This distance $d$ is called the `Robinson--Foulds metric'; it satisfies the triangle inequality and it can be computed quickly.

Turning to the consensus question, given a sequence of rooted phylogenetic $X$-trees $T_1, T_2, \ldots, T_k$, let $H^*$ be the set of clusters that are present in
more than half of the corresponding hierarchies. In other words, if $H_i$ is the hierarchy on $X$ corresponding to $T_i$ then 
$H^* = \{C \in \bigcup_{i=1}^k H_i:   |\{j: C \in H_j | > k/2\}.$  The following lemma shows that $H^*$ forms the set of clusters of a tree, the so-called  `majority rule' consensus tree.  

\begin{lemma}
$H^*$ forms a hierarchy, and so corresponds to a rooted phylogenetic $X$-tree.
\end{lemma}
{\em Proof: }
Suppose $C, C' \in H^*$. By the `pigeonhole principle',  there must be some hierarchy $H_j$ that contains both  $C$ and $C'$. Consequently, $C$ and $C'$ are either disjoint or one is nested in the other. As this holds for all $C,C' \in H^*$, condition {\bf H1} holds.  Moreover, {\bf H2} holds also, since $\{x\}$ and $X$ are elements of $H_j$ for every $j$ and every $x\in X$.  Thus, $H^*$ forms a hierarchy and so (by Lemma~\ref{equiv}), corresponds to a rooted phylogenetic $X$-tree that is unique up to equivalence.
\hfill$\Box$

\bigskip 
The majority rule consensus tree has the nice combinatorial property that it comes as `close as possible',  on average,  to the input trees $T_1, T_2, \ldots, T_k$ under the Robinson-Foulds metric; more precisely,  it is a `median' tree $T$ that  minimizes $\sum_{i=1}^k d(T, T_i)$ \cite{day}.

\subsection{Unrooted phylogenetic trees.}
Rooted trees appeal to biologists since they show evolution happening in the time, from the past to the present. But it is often more convenient to consider unrooted trees. One reason is that most methods for building trees from data can usually do so only up to the placement of the root, and so produce unrooted trees (figuring out where the root goes usually comes later).  Also, from a mathematical perspective, unrooted trees are often the more natural object to consider.  The choice to work with either
rooted or unrooted trees is somewhat analogous to the distinction in classical geometry between the affine and projective settings  (respectively), where also  one viewpoint may have advantages over the other,  depending on the questions at hand.

\bigskip

{\bf Definition:} An {\em (unrooted) phylogenetic $X$-tree} is a tree $T$ with leaf set $X$ and with every interior (i.e. non-leaf) vertex of degree at least two.  If the degree of
every non-leaf vertex is exactly three, we say that $T$ is a {\em binary phylogenetic $X$-tree}.

\bigskip

Here is  a first property of such trees, which will be useful in the next section.
\begin{lemma}
\label{edge}
Any unrooted binary phylogenetic tree $T$ with $n$ leaves has $2n-3$ edges.
\end{lemma}
{\em Proof:}  
A standard result in elementary graph theory states that a  connected graph is a tree if and only if the number $N$ of vertices exceeds the number $E$ of edges by 1.
So if  our tree $T$ has $i$ interior vertices, we have $N=i+n$ and so:
\begin{equation}
E=i+n-1.
\label{Eeq}
\end{equation}
Also, for any graph, the `handshake lemma' tells us that the sum of the degrees of the vertices of any finite graph equals $2E$, since each edge is counted twice in this sum.
Now, for our tree $T$,  the sum of the degrees is $[1+ 1+ \cdots +1 (n \mbox{   times})] + [3+ 3+\cdots + 3 (i \mbox{ times})]$ and so: 
\begin{equation}
\label{2Eeq}
2E= n + 3i.
\end{equation}
Combining Eqns. (\ref{Eeq}) and (\ref{2Eeq}) we see that $i=n-2$, and so $N = i+n = 2n-2$, which implies that $E=N-1=2n-3$. This completes the proof of Lemma~\ref{edge}.
\hfill$\Box$

\bigskip
{\bf Notation:}  We will let $R(X)$ and $U(X)$ be the sets of rooted and unrooted phylogenetic $X$-trees (up to equivalence), and
$RB(X)$ and $UB(X)$ will denote the sets of rooted and unrooted \underline{binary} phylogenetic $X$-trees.
Thus when $X$ has just four elements, $UB(X)$ consists of the three  {\em quartet trees}, while $U(X)$ has one additional `star' tree, that has a single non-leaf vertex of degree four.   

When $X=[n]=\{1,\ldots, n\}$, we will write $R(n), U(n), RB(n)$ and $UB(n)$ for $R(X), U(X), RB(X)$ and $UB(X)$, respectively.

  \subsection{Unrooting and counting trees.}
  \label{unro}

Counting trees has a long tradition in mathematics, with Cayley's $n^{n-2}$ formula from 1889 for the total number of trees on $n$ labelled vertices,  the most famous example.
Counting binary phylogenetic trees turns out to be much easier, and it has a history that dates back to even earlier mathematical work, contemporary with Darwin \cite{sch}.   To  explain this, we first describe a close connection between rooted and unrooted phylogenetic trees. There are two natural ways to associate an unrooted phylogenetic $X$-tree with a rooted   tree.
\begin{itemize}
\item[] {\em Adding an outgroup:}  Take a rooted phylogenetic tree on $X-\{x\}$ and attach $x$ to the root of $T$ by a new edge. Species $x$ is called an `outgroup' species.
\item[] {\em Suppressing the root:}  Simply ignore the root vertex $\rho$; if it has degree 2 then delete it and identify its two incident edges, while if the root has degree at least three
then just treat this vertex as an interior vertex with no special root status.   An example is shown in Fig.~\ref{tree1}.
\end{itemize}

\noindent Notice that the operation  `Adding an outgroup' provides a bijection: $$o: R(X-\{x\}) \rightarrow U(X)$$ which restricts to a bijection from $RB(X-\{x\})$ to $UB(X)$.
On the other hand, `suppressing the root' results in a surjective map:  $$s: R(X)  \rightarrow U(X)$$ which restricts to a surjective map from $RB(X)$ to $UB(X)$. Moreover, the number
of elements of $RB(X)$ which map to the same tree in $T \in UB(X)$ is the number of edges in $T$, which Lemma~\ref{edge} tells us is $2n-3$. ($n=|X|$).
These observations show us that:  $$|RB(X)| = (2n-3)|UB(X)|= (2n-3)|RB(X-\{x\}|.$$
In particular, if $r(n)$ is the number of rooted binary phylogenetic trees on a leaf set of size $n$ then
$r(n) = (2n-3)r(n-1)$, which, together with $r(2)=1$, gives:
$$r(n) = 1 \times 3 \times 5  \times \cdots \times (2n-3).$$
This product of the odd numbers is often written as the double factorial (in this case, $(2n-3)!!$). Notice that it can be expressed in terms of ordinary factorials and powers of $2$ as follows:

\begin{equation}
\label{rntree}
r(n) = \frac{(2n-2)!}{(n-1)!2^{n-1}}.
\end{equation}
Graph theorists may recognize this quantity: it is  the number of perfect matchings of a complete graph on $2n-2$ vertices. In other words, if there are $2n-2$ people in a room, Eqn.  (\ref{rntree}) counts the number handshake scenarios in which each 
 person shakes hands with precisely one other person.   The bijection between this set of scenarios and set $RB(n)$ is an interesting but nontrivial exercise \cite{dia}.

Applying Stirling's approximation $n! \sim \sqrt{2\pi}\cdot n^{n+\frac{1}{2}} e^{-n}$ to Eqn. (\ref{rntree}), reveals  that $r(n)$ grows very rapidly. For example, $r(10)$ is around than 34 million, while $r(30$) is more than $10^{38}$.
Biologists often want to build trees for hundreds (or even thousands) of species; so it's no surprise  that mathematics has an important role to play in this task, as it would be impossible to check each tree to see how well it might `fit the data'.

There is another way to arrive at Eqn (\ref{rntree}), by using generating functions.
If we consider the formal power series $\phi(x) =x +  \sum_{n \geq 2} r(n)  \frac{x^n}{n!}$ then:
$$\phi(x) = \frac{1}{2}\phi(x)^2 + x,$$
since deleting the root of a tree $T \in RB(n)$ for $n>2$ results in two rooted binary trees (or an isolated leaf) on leaf sets $Y_1$ and $Y_2$ that partition $[n]$.  
Solving this quadratic equation gives $\phi(x) = 1-\sqrt{1-2x}$, from which $r(n)$ pops out as $n!$ times the coefficient of $x^n$ in $\sqrt{1-2x}$.
While this is a more complicated derivation, generating functions  turn out to be very useful in other applications -- for example, in deriving exact explicit formulae for the number of `forests' of rooted binary trees on a given leaf set.

For the number $u(n)$ of unrooted binary trees on a leaf set of size $n$, the bijection $o$ described above gives $u(n) = r(n-1) = (2n-5)!!$
Nonbinary phylogenetic trees (rooted and unrooted) can also be counted using recursions, but a closed-form expression like that for binary trees is lacking.

\section{Tree shapes.}
  
\begin{figure}[htb]
\centering
\includegraphics[scale=1.0]{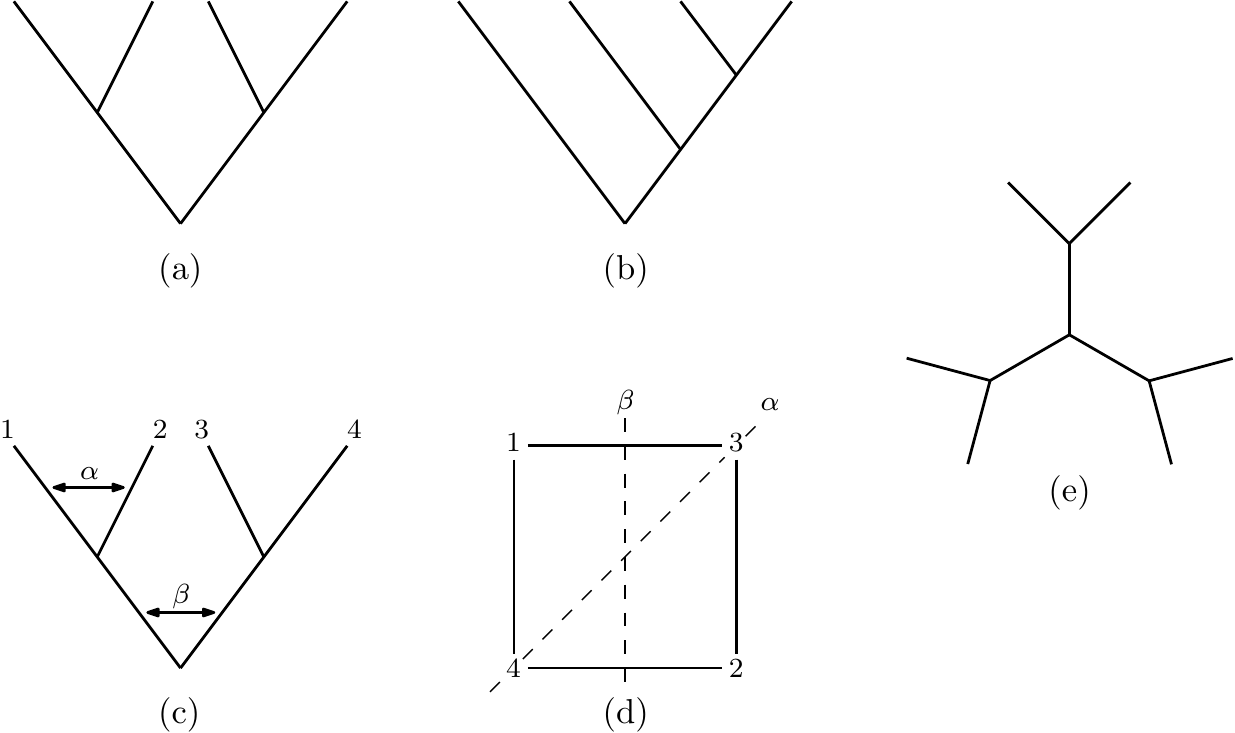}
\caption{The two tree shapes for rooted binary trees on four leaves:  (a) the `fork' and (b) the `pectinate' tree shape.  The stabilizer subgroup of a phylogenetic tree having the fork shape (c) corresponds to the dihedral group of symmetries of a square (d). The two symmetries shown ($\alpha$ and $ \beta$ in (c)) correspond to reflections. In (e) an unrooted tree shape with a symmetry of order $3!$ about the central vertex is shown.}
\label{shapesfig}
\end{figure}

If we ignore the labeling of the leaves of a rooted or unrooted phylogenetic tree,  we obtain a `tree shape'. For example, when $n=4$, there are two rooted binary tree shapes: the `fork' tree shape and the `pectinate' tree shape, shown in Fig.~\ref{shapesfig}(a, b).   Biologists are interested in the shapes of trees, since they shed light on the process of speciation and extinction in evolution.  

Elementary group theory provides a nice trick to count the number of phylogenetic $X$-trees of a given shape using the `orbit-stablizer theorem'. Given a finite
group $G$ which acts on a set $S$, let $O(s) = \{g\cdot s: g \in G\} \subseteq S$ denote the orbit of $s$ under the action of $G$, and let ${\rm Stab}(s) = \{g \in G: g\cdot s = s\} \subseteq G$ be the stabilizer subgroup of $G$. Then the orbit stabilizer theorem provides a bijection between the orbit of $s$ and the cosets of ${\rm Stab}(s)$ in $G$, and so, in particular:
\begin{equation}
\label{OSeq}
|O(s )| = |G|/|{\rm Stab}(s)|.
\end{equation}

There is a natural action of the symmetric group $\Sigma_n$  of permutations on $[n]$ on the set $R(n)$: given $\sigma \in \Sigma_n$, simply permute the leaves of each tree $T$ by replacing leaf $x$ by leaf $\sigma(x)$.  This action restricts to an action on the set $RB(n)$ of rooted binary trees, and so, by (\ref{OSeq}), the number of trees in $RB(n)$ that have the same shape as some tree $T$ is  $n!/|{\rm Stab}(T)|$.  Now ${\rm Stab}(T)$ is a group of order $2^{s(T)}$ where $s(T)$ is the number of {\em symmetry vertices} of $T$ -- these are interior vertices for which the two subtrees of $T$ that the vertex separates from the root have the same shape.  For example, for a phylogenetic tree having the `fork' tree shape in Fig.~\ref{shapesfig}(a),  $T$ has three symmetry vertices and so  ${\rm Stab}(T)$ is a group of order eight. This group turns out to be 
isomorphic to the dihedral  group of rotational and reflectional symmetries of a square, as illustrated in Fig.~\ref{shapesfig}(c,d). In particular, for any set $X$ of size four,  there are precisely $4!/2^3 = 3$ rooted binary $X$-trees that have the  shape of the fork tree; by  contrast,  the pectinate tree (Fig.~\ref{shapesfig}(b)) has only one symmetry vertex, and so there are 12 rooted binary phylogenetic $X$-trees of this shape.

For unrooted binary trees and nonbinary trees, similar formulae apply, though more complex symmetries arise; for example,  an unrooted binary tree can have a 2-fold symmetry about a central edge  and, in the case of the tree shape shown in Fig.~\ref{shapesfig}(e), a  symmetry of order $3!$ about a central vertex.  In general,  the group ${\rm Stab}(T)$
 is usually described using `wreath products'.  
 

\subsection{The shape of evolving trees, and  the many roads that lead to the Yule--Harding distribution}
\begin{figure}[htb]
\centering
\includegraphics[scale=1.0]{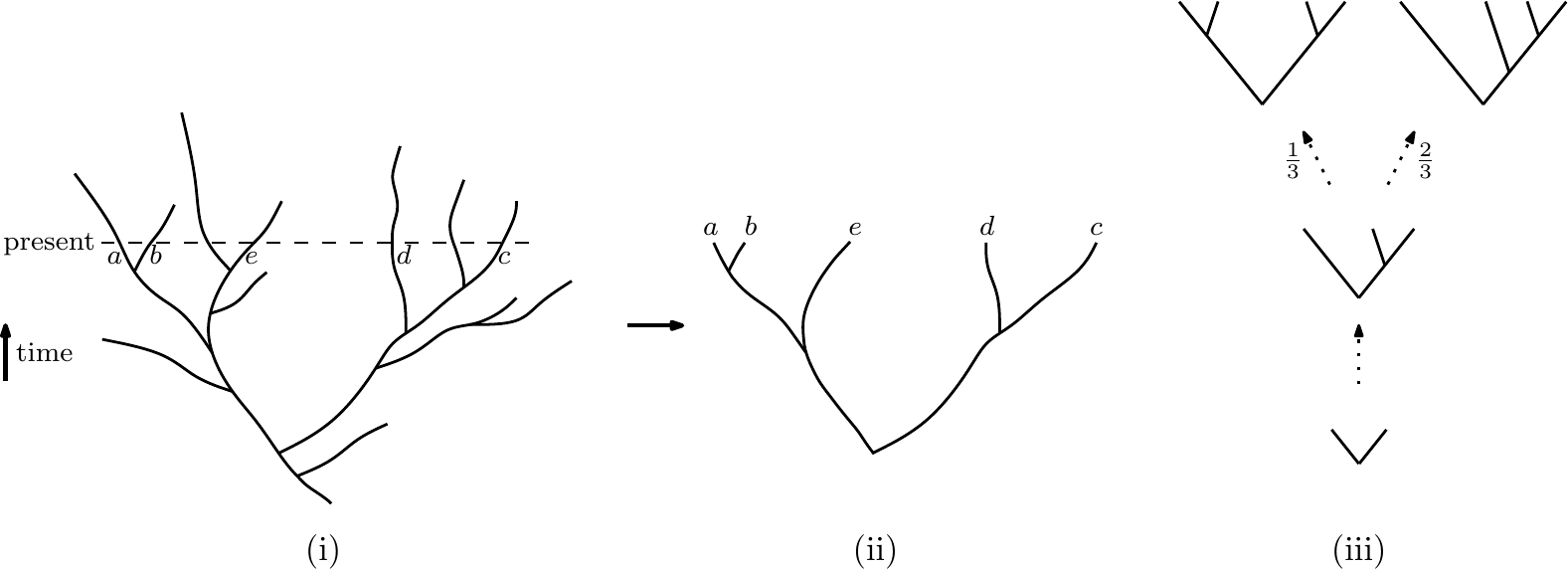}
\caption{(i)  A birth-death tree showing speciation and extinction. (ii) The associated `reconstructed tree'. (iii) Growing a tree by the YH process.}
\label{birth}
\end{figure}

Although extinction has played a major role in the history of life (after all, most species are extinct),  suppose we sample some subset $X$ of species present today (species $a-e$ in Fig.~\ref{birth} (i)) and then consider the minimal  tree linking  these species. This results in  the so-called `reconstructed tree'  illustrated in Fig.~\ref{birth} (ii).   Let's think of this as a rooted phylogenetic $X$-tree (ignoring the length of the edges).   It turns out that, under very general assumptions concerning the speciation-extinction process, many models predict an identical and simple discrete probability distribution on $RB(X)$ \cite{lam}.      

This distribution is called the  {\em Yule--Harding (YH)} distribution, and it is easily described as follows:   To obtain a binary tree shape, we start with a tree shape on two leaves and sequentially attach leaves -- at each step attaching a new leaf to one of the leaf edges chosen uniformly at random from the tree constructed so far.  For example, the probabilities of generating the fork and pectinate tree shapes are 1/3 and 2/3, respectively, since from the (unique) tree shape on three leaves, we can place a new leaf on one leaf-edges to obtain a fork tree shape, or on two edges to obtain a pectinate tree shape (see Fig.~\ref{birth}(iii)).  

Once we have built up a tree with $n$ leaves in this way we obtain a random tree shape on $n$ leaves, and we can now label the leaves of this tree shape according to a permutation on $\{1,2, \ldots, n\}$ chosen uniformly at random. This is the Yule--Harding probability distribution on $RB(n)$. 
   Curiously, a quite different process that arises in  population genetics, and which proceeds backward in time (rather than forward, like Fig.~\ref{birth}(iii)) also leads to the YH distribution when we ignore the length of the edges. This is the celebrated `coalescent process' of Sir John Kingman from the early 1970s.  
 
We now explain how to compute the probability of a YH tree shape and of any rooted phylogenetic tree with this shape.  First  let's grow a  tree under the YH process until it has $n$ leaves, and then select one of the two subtrees incident with the root (say the `left-hand one', since the orientation in the plane plays no role) and let $Z_n$ denote the number of leaves in this tree.  Remarkably, $Z_n$ has a completely flat distribution.
\begin{lemma}
\label{unif_lem}
$Z_n$ has a uniform distribution between 1 and $n-1$, so
$$\PP(Z_n = i) = \frac{1}{n-1}, \mbox{ for } i = 1, \ldots, n-1.$$
\end{lemma}

{\em Proof:} 
The random process $Z_1, Z_2, \ldots $ can be exactly described as a special case of a classical process in probability called  {\em Polya's Urn}. This consists of an urn that initially has $a$ blue balls and $b$ red balls.  At each step, a ball is sampled
uniformly at random and it is returned to the urn along with another ball of that same color. In our setting, $a=b=1$ and `blue' corresponds to the left-hand subtree and `red' the right-hand subtree in the Yule--Harding tree.  At each step the uniform process of  leaf attachment ensures that $Z_n$ has exactly the same probability distribution as the number of blue balls in the urn after $n-2$ steps. It is well known, and easily shown by induction, that in Polya's Urn with $a=b=1$, the proportion of blue balls has a uniform distribution.  

\hfill$\Box$

This lemma provides the key to computing the YH probability of a tree exactly, as follows.
\begin{proposition}
\label{YHprob}
For any particular tree $T \in RB(n)$, the probability $\PP_{YH}(T)$ of generating $T$ under the YH model is given by:
$$\PP_{YH}(T) = \frac{2^{n-1}}{n!\prod_{v \in I(T)} \lambda_v}$$
where $I(T)$ is the set of interior vertices of $T$, and where $\lambda_v$ is $-1$ plus the number of leaves of $T$ that are descendants of $v$.
\end{proposition} 
For example, for the tree in Fig.~\ref{tree1}(a),  $\PP_{YH}(T) = \frac{2^4}{5!\times 4\times 3\times 1^2}=\frac{1}{90}$, while for Fig.~\ref{birth}(ii),  $\PP_{YH}(T) =\frac{1}{60}$.

\medskip

{\em Proof:} 
Suppose that the two maximal subtrees $T_1$ and $T_2$ of $T$ are of size $k$ and $n-k$, where we may assume that $2k \leq n$.
By Lemma~\ref{unif_lem} the probability of such a size distribution is $2/(n-1)$ if $2k<n$ and $1/(n-1)$ if $2k=n$.
Conditional on this division, the number of ways to select leaf sets for $T_1$ and $T_2$ that partition $[n]$ is
$\binom{n}{k}$ when $2k<n$,  and
$\frac{1}{2}\binom{n}{k}$ when $2k=n$ (the factor of $\frac{1}{2}$ recognises that the order of $T_1$ and $T_2$ is interchangeable in $T$ when they have the same number of leaves).
 By  the Markovian nature of the YH process, each of these two subtrees also follows the YH distribution. This leads to
the recursion:
$$\PP(T) = \frac{2}{n-1} \binom{n}{k}^{-1} \PP(T_1)\PP(T_2),$$
from which Proposition (\ref{YHprob}) now follows by induction.  
\hfill$\Box$

\medskip

Notice that the YH process leads to a different probability distribution on $RB(n)$ from that obtained by  simply selecting a tree uniformly at random from $RB(n)$, which would assign each $T \in RB(n)$ the probability  $1/(2n-3)!!$  To see that this is different from the YH distribution, observe that  the probability of obtaining a tree with the `fork' in Fig.~\ref{shapesfig}(a)  has probability $\frac{1}{5}$ under a uniform distribution on $RB(4)$ (since only 3 of the 15 trees in $RB(4)$ have that shape) and $\frac{1}{3}$  under YH.   There are several more general differences between the uniform and YH distributions. For instance,  in YH trees on $n$ leaves, the expected average number of edges between the root and the leaves grows at the rate $\log(n)$, while for uniform binary trees, it grows at the rate $\sqrt{n}$.   YH trees also tend to be more `balanced' than uniform trees, where balance refers to the average  difference between the sizes of the two daughter subtrees in the tree, as one ranges over the interior vertices of the tree.  For example, Lemma~\ref{YHprob} shows that the probability that tree with $n$ leaves  generated under the YH distribution has a single leaf adjacent to the root is $\frac{2}{n-1}$, while for the uniform distribution the corresponding probability is $\binom{n}{1} \frac{r(n-1)}{r(n)} =\frac{n}{2n-3}$ which converges to $\frac{1}{2}$ as $n$ grows.  It turns out that many `real' phylogenetic trees tend to have a degree of balance somewhere between that predicted by the YH and uniform distributions;  explaining why has required some new mathematical and statistical insights \cite{ald, lam}.

\section{Trees, splits and character data}
\label{splitsec}

In Section~\ref{unro} we described a one-to-one correspondence between unrooted phylogenetic $X$-trees and rooted phylogenetic $X$-trees on $X-\{x\}$ (for any $x \in X$), and thereby
to hierarchies on $X-\{x\}$.  However, the choice of a particular element $x \in X$ is completely arbitrary, so we  seek a  more satisfactory way to describe an unrooted phylogenetic $X$-tree.  This is based on  the notion of an {\em $X$-split}, which is a bipartition of $X$ into two nonempty parts ($A$ and $B$, say), and written as $A|B$.  Such a notion has clear biological meaning -- for example, we can divide all life into the `vertebrates' and the `invertebrates'.  
Given any phylogenetic $X$-tree $T$, if we delete any particular edge $e$ of $T$  and consider the leaf sets of the two connected components of the resulting disconnected graph we obtain a corresponding $X$-split, which we will refer to as a {\em split of $T$} that corresponds to $e$.    For example, for each $x \in X$, every phylogenetic $X$-tree has the {\em trivial split} $\{x\}|X-\{x\}$,  corresponding to the edge incident with leaf $x$. 

Notice that any two splits  $A|A'$ and $B|B'$ of the same phylogenetic $X$-tree have the property that one of the four intersections $A \cap B, A \cap B', A' \cap B, A' \cap B'$ is empty. 
If a collection  $\Sigma$ of $X$-splits has this property, we say that $\Sigma$ is  {\em pairwise compatible}.  This is the unrooted analogue of the hieararchy property {\bf H1}, so it is not surprising  that Lemma~\ref{equiv} has an equivalent formulation for unrooted trees:

\begin{quote}
A collection $\Sigma$ of $X$-splits is the set of splits of some unrooted phylogenetic $X$-tree $T$ if and only if 
$\Sigma$ is pairwise compatible and contains the trivial splits.   Moreover, $T$ is uniquely determined up to equivalence by $\Sigma$.
\end{quote}
We can extend the notion  of splits further. Instead of deleting a single edge, we may delete a set $E'$ of $k \geq 1$ edges from a tree, and consider the resulting $k+1$ components of the disconnected graph $T-E$. This gives an equivalence relation $\sim$ on $X$ where $x \sim y$ precisely if $x$ and $y$ are connected in $T-E$.  The equivalence classes of $\sim$ comprise a partition of $X$ into at most $k+1$ parts.  Any such partition of $X$ that can be obtained in this way is said to be {\em convex} on $T$, a concept that is relevant to the next part of the story.

\subsection{Characters, homoplasy and a `perfect phylogeny'}
\label{carsec}

A function from the set of species $X$ into some set $S$ of $r$ states is referred to by biologists as  an $r$-state {\em character} on $X$.   For example, $f(x)$ might be a morphological character that describes the number of legs that species $x$ has, or a genetic character the describes the nucleotide at a particular position in a genetic sequence for species $x$.  That is, $f(x)$ describes some `characteristic' of $x$ that we compare across other species in $X$.
A hypothetical example of four characters across a set of eight well-known species, is provided in the table  below, and will serve to illustrate several ideas that follow.  If we regard each of the possible states as (say) letters of the alphabet, then we can associate a four-letter `word' to each species.  

\begin{table}[h]
\begin{center}
\begin{tabular}{llllll}
\hline
Species  &     Character &1  & 2 & 3 & 4 \\
\hline
Kangaroo & &  T & R & U & E \\
Chimpanzee & & B & R & E & T  \\
Human & & B & R &O & E\\
Gorilla & & C & O & E &E \\
Hippopotamus & & C & A & P & O \\
Whale & & C & A & U & P \\
Lion & & D & R & A & O  \\
Tiger & & D & R & U & G \\
\end{tabular}
\end{center}
\end{table}

If a phylogenetic $X$-tree describes the evolution of a set of species, a character tells us  the states of the species at the leaves, but not  of the hypothetical ancestral species that correspond to the interior vertices of the tree.  There are myriad ways to explain how the character could have evolved in the tree from some ancestral state at the root.  It is possible that in a  path from the root to a leaf a {\em reversal}  occurs, where a state $s_1$ changes to state $s_2$  and later back to $s_1$; for example, in birds wings first evolved and then in some species (e.g. kiwi) disappeared again. It is also possible for `convergent' evolution' to occur, where state $s_1$ at some vertex changes to $s_2$ down two edge-disjoint paths that start from that vertex.
Again wings provide an example - from an ancestor of birds and mammals, wings evolved both in birds and in mammalian bats.  A character whose evolution on a given tree can be explained without postulating any reversal or convergent events is said to be 
 `homoplasy-free'.    Homoplasy-free evolution might be expected to hold when the number of potential states is very large, so each change is likely to  be to a new state (for example,  the order of genes on a chromosome under random rearrangement operations); or when the rate of state change is very low.

The notion of homoplasy-free can be defined more easily if we suppress the rooting of the tree and so consider unrooted trees.  Formally, we will say that a character $f$ on $X$ is {\em homoplasy-free} on an unrooted phylogenetic $X$-tree $T$ if states can be assigned to the interior vertices of $T$ so that  the path between any two vertices that are assigned the same state contains only vertices that are also assigned the same state.   In other words, $f:X \rightarrow S$ has an extension $F: V\rightarrow S$ to the set $V$ of all vertices of $T$  so that for each $\alpha \in f(X)$,  the subgraph of $T$ induced by the set of vertices $v$ with $F(v)=\alpha$ is connected.  There are two other ways to characterize when a character $f:X \rightarrow S$ is homoplasy-free on a phylogenetic $X$-tree $T$:
\begin{itemize}
\item  $f$ has an an extension $F$ to all the vertices of $T$ for which $F$ assigns different states to the endpoints of $|f(X)|-1$ edges (equivalently, at most $|f(X)|-1$  edges);
\item  the  partition of $X$ induced by the equivalence relation ``$x\approx x' \Leftrightarrow f(x) = f(x')$''  is convex on $T$  (as defined just prior to Section~\ref{carsec}).
\end{itemize}
Notice that homoplasy-free is considerably weaker than requiring that the actual evolution of the character on some rooting of the tree involved no reversals or 
convergent evolution --- it merely requires that the character {\em could} have evolved in this way.

A sequence $(f_1, f_2, \ldots, f_k)$ of characters on $X$ is said to have  a {\em  perfect phylogeny}  if and only if there exists a phylogenetic $X$-tree  on which each character is homoplasy-free (the tree is said to be a perfect phylogeny for those characters).   We will see soon that our eight-species example above forms a perfect phylogeny.

The computational problem of  determining whether or not a collection of characters has a perfect phylogeny is NP-complete in general, but a polynomial-time algorithm exists  when a bound is placed on either the number of characters or the
number of states per character.  In the special cases where $r=2$ and $r=3$ a collection of $r$-state characters has a perfect phylogeny if and only if every subset of the characters of size $r$ has one.  However the `if' direction fails for larger values of $r$, as there is a set of $\lfloor{\frac{r}{2}}\rfloor \cdot \lceil{\frac{r}{2}}\rceil +1$ characters on $r\geq 4$ states that do not have a perfect phylogeny even though every proper subset does \cite{shu}.   The existence of a perfect phylogeny for a sequence  of characters on $X$ also has an attractive graph theoretic characterization involving intersection graphs (for details, see \cite{sem}; more recent graph-based analysis of related approaches appears in \cite{bon}). 

When a sequence of characters has a perfect phylogeny $T$, we can also ask when it is unique.  A necessary condition for this   is that $T$ is binary;  otherwise,  we could arbitrarily resolve any vertex of $T$ of degree greater than three, and obtain a different tree on which all the characters were homoplasy-free.   
An interesting question now arises: what is the smallest number $h(n)$ so that for each $T \in UB(n)$, there is a sequence of $h(n)$ characters on $[n]$ that has $T$ as a unique
perfect phylogeny?  If we restrict ourselves to binary characters, then $$h(n) =  n-3,$$ 
since for any $T \in UB(n)$, the sequence of  2-state characters corresponding to the $n-3$ non-trivial splits of $T$  have this tree as their unique perfect phylogeny, and none of these characters could be removed (otherwise we could contract the corresponding edge and still obtain a tree on which the characters were homoplasy-free). But  what if we do not insist on restricting ourselves to 2-state characters, or $r$-state characters for any fixed $r$.  Is it possible that
$h(n)$ might grow more slowly than linearly with $n$; perhaps $\sqrt{n}$ or even
$\log(n)$ characters might suffice?   Surprisingly, it  turns out that $h(n)$ is never more than four.

\begin{theorem}[Four characters suffice]
\label{amaze}
For any binary phylogenetic $X$-tree $T$,  there is a set $S_T$ of at most four characters for which $T$ is the {\em only} perfect phylogeny.
\end{theorem}

\begin{figure}[htb]
\centering
\includegraphics[scale=0.9]{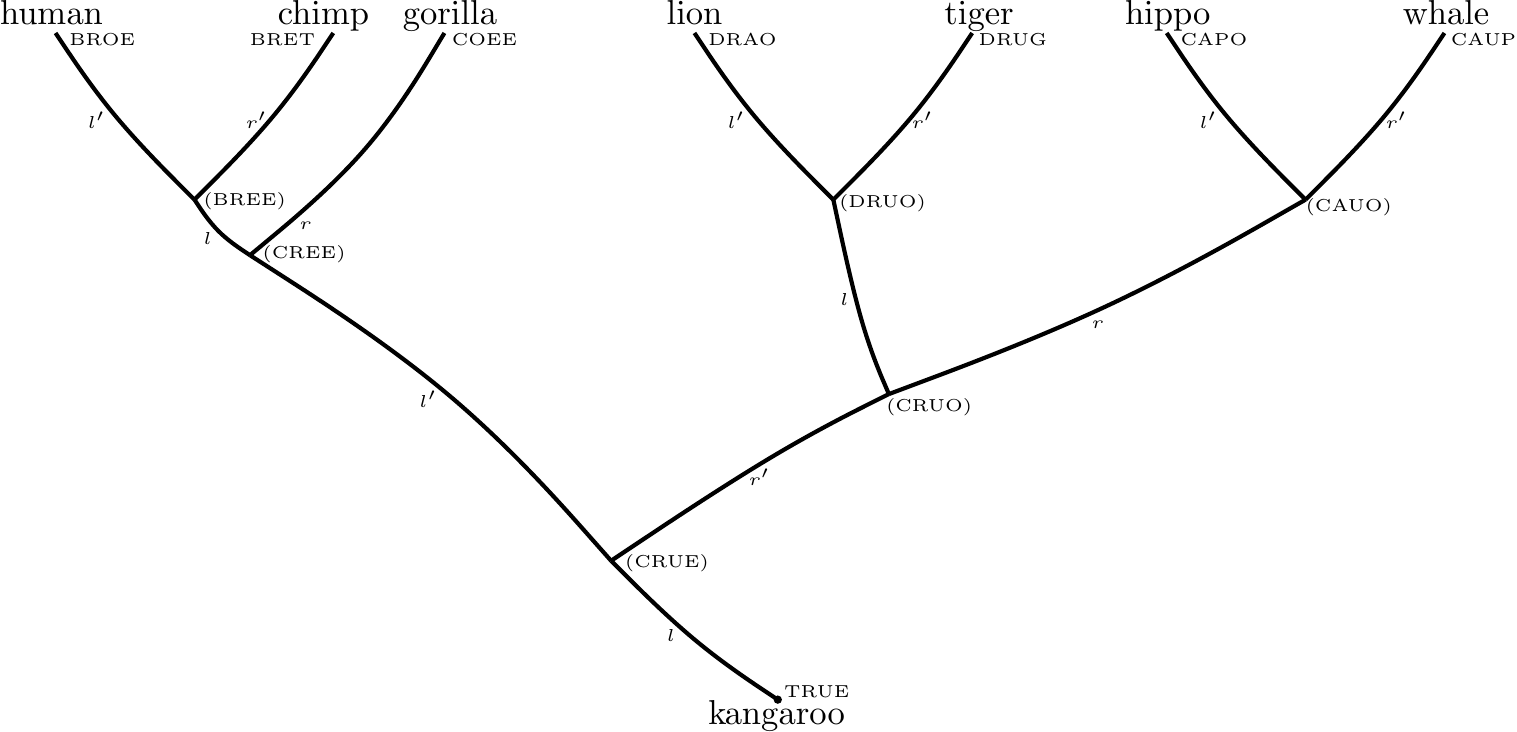}
\caption{The unique perfect phylogeny $T$ for the four characters described in the table above.  The assignment of ancestral states (in brackets) is also uniquely determined.}
\label{tree4}
\end{figure}

An example of $S_T$ is provided by the four hypothetical characters described for the eight species in the table above.  It is easily seen that the tree $T$ shown in
Fig.~\ref{tree4} is a perfect phylogeny for this data set (this tree, incidentally,  is the one biologists generally accept).  But what can also be shown is that $T$ is the only such perfect phylogeny for these four characters; moreover the states at the interior vertices (shown in brackets) are uniquely determined by the homoplasy-free condition.   

A recipe to generate a set  $S_T$ is indicated by the letters $l, r, l', r'$ on the edges of the tree. These correspond to alternating `left' $(l, l')$ and `right' $(r, r')$ orientations as one moves up the tree, under an arbitrary planar embedding.  Now, suppose any edge on which $l$ is places causes a state change for the first character, any edge on which $r$ is placed causes a state change for the second character, and similarly any edge on which $l'$ (resp. $r'$) is placed causes a state change for the third (reap. fourth) character.  State changes are always  to a new state for that character (to ensure the homoplasy-free condition;  a state present in one character is free to re-appear in a different one).  For example, the bottom-most $l$ causes TRUE to change to CRUE. By  following this procedure for any binary tree on any number of leaves it can be shown that $S_T$ satisfies Theorem \ref{amaze} (for further details, see \cite{hub}).

\subsection{When the data are not `perfect'.}

The homoplasy-free condition is very strong.  A natural relaxation of it, given a character $f$  and a phylogenetic tree $T$,  is to score $T$ by the smallest number of edges of $T$ which need to have differently assigned states at their endpoints in order to extend $f$ to all the vertices of $T$.  This score is called the {\em parsimony score} of the character on $T$, denoted, $ps(f, T)$. By the equivalent description of the homoplasy-free condition above, we have  $ps(f,T) \geq |f(X)|-1$, with equality if and only if $f$ is homoplasy-free.

Since there are exponentially many extensions $F$ of $f$ to $T$ it might be suspected that computing $ps(f,T)$ is hard. However, in 1971, biologist Walter Fitch proposed a fast algorithm, which was formally verified by mathematician John Hartigan in 1973.  This `Fitch--Hartigan algorithm' proceeds via a dynamical programming approach, and it also provides an explicit extension $F$ that minimises the number of state changes in the tree.  For 2-state characters, there is also a nice connection between the parsimony score and  Menger's min-max theorem in graph theory.

Given a sequence of characters on $X$, a {\em maximum parsimony tree} for this data is a phylogenetic tree $T$  that minimizes the sum of the parsimony scores  of the characters.  
Finding such a tree can be phrased as a `Steiner tree' problem in a sequence space, and it  turns out to be NP-hard,  though branch and bound algorithms exist.

We saw that no sequence of 2-state characters shorter than linear in $n$ can give a unique perfect phylogeny. But what if we just want a unique most parsimonious tree?  That is, for each tree $T \in UB(n)$, is there a sequence $\eta(T)$ of  2-state characters of length $k=k(n)$ that is sub-linear in $n$ and for which $T$ is the unique most-parsimonious tree? A  simple counting argument sets an absolute lower bound on $k$. Let $S(n,k)$ be the set of sequences of 2-state characters on $[n]$ of length $k$.  Then $k$ must be at least large enough for the function $T \mapsto \eta(T)$ from $UB(n)$ to $S(n,k)$ to be one-to-one. Since $S(n,k) =2^{nk}$ this requires that  $|UB(n)| \leq 2^{nk}$, which can be rewritten as $k \geq \frac{1}{n}\log_2|UB(n)|$.  If we now invoke  Eqn. (\ref{rntree}) and Stirling's approximation for $n!$ to calculate $|UB(n)|$ we see that  $k$ must grow at least at  the rate $\log(n)$.   Remarkably,  it was recently shown \cite{cha} that this primitive growth rate is achieved, and by a function $\eta$ that can be constructively implemented. Moreover, the homoplasy score per character of the resulting sequences $\eta(T)$ on $T$ necessarily tends to infinity as $n$ grows, so this encoding is very `far' from supporting a perfect phylogeny.

We will return to maximum parsimony in Section~\ref{markovsec}.

\section{Metric properties of trees.}
\label{met}

So far, we have regarded the edges of our trees as being unweighted; however it is useful -- both in biology and in mathematics --  to assign weights or lengths to the edges (often called `branch lengths' in biology). For instance the length of an edge could correspond to
evolutionary time, or some measure of the amount of genetic change along that edge. Assigning lengths to edges brings in a further tool to help study and reconstruct trees.

Firstly, notice that if we have a phylogenetic $X$-tree $T$, and some function $w$ that assigns strictly positive weights to each edge of the tree, then we can define a metric $d = d_{(T,w)}$ on $X$ by
letting $d(x,y)$ be the sum of the weights of the edges on the path in $T$ connecting $x$ and $y$. When $d$ can be represented by a tree in this way, we say it has a {\em tree representation} (on $T$). 

This leads to two natural questions: 
\begin{itemize}
\item Does every metric on $X$ have a tree representation? 
\item Is the choice of $T$ and $w$ in a tree representation unique? 
\end{itemize}  
The answers to these questions are `no' and `yes' respectively. Let's consider the first question.  

\begin{figure}[htb]
\centering
\includegraphics[scale=1.2]{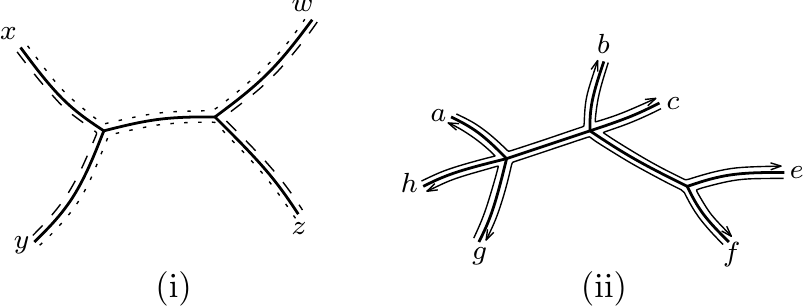}
\caption{(i): Here $d(x,y)+d(w,z)$ is smaller than $d(x,w)+d(y,z)$ (which, in turn, equals $d(x,z)+d(y,w)$); (ii) a tour of the tree that covers every edge exactly twice.}
\label{tree5}
\end{figure}

When $|X|=3$, it is an  easy exercise to show that every metric $d$ on $X$ can be represented as a tree metric.  But this result is particular to $|X|=3$ and already  runs into problems when $|X|=4$. It is instructive to see why. Consider the three pairwise sums:
$$d(x,y) + d(w,z), \mbox{  }   \mbox{  }  d(x,z) + d(y,w), \mbox{   }    \mbox{  }  d(x,w)+ d(y,z).$$
If $d$ has a tree representation ($d=d_{(T,w)}$), then two of these pairwise sums must be equal, and larger than the third, regardless of the choice of $T$.   This is illustrated in Fig.~\ref{tree5}(i).
This  `four-point condition' is not usually satisfied by an arbitrary metric $d$ on a set of size four, but when it is, it turns out that $d$ can be represented on a tree.
What is much more remarkable is that, for any $X$, the four point condition holds for all subsets of $X$ of size 4 if {\em and only if} $d$ has a tree representation.
This result, in various forms, dates back to the 1960s and has been rediscovered several times. 

Consider now the second question: the uniqueness of a tree representation.  As before, this question was resolved many decades ago, and  uniqueness of both the unrooted tree and
the strictly positive edge weights holds; in other words, for trees $T, T' \in U(X)$, and $w,w'$ strictly positive, we have:
\begin{equation}
\label{unique}
d_{(T,w)} = d_{(T',w')} \Longrightarrow T=T' \mbox{ and } w=w'.
\end{equation}
Moreover, to reconstruct a phylogenetic tree with $n$ leaves we do not usually need all the $\binom{n}{2}$ possible $d$-values; for a binary tree $T$,  a subset of $2n-3$ carefully chosen pairs of elements from $[n]$  suffice to uniquely determine both  $T$ and $w$ from the value of $d_{(T,w)}$ for those pairs.

A variety of 
fast  (polynomial-time) methods have been devised for building a phylogenetic $X$-tree from an arbitrary distance function $d$ on $X$, the most popular being called `Neighbor Joining'.  A desirable property of such methods is that when a distance function has a tree representation then the method will return the underlying tree.  Moreover,  several such methods (including Neighbor Joining) possess a provable  `safety radius' $\epsilon$ around $d=d_{(T,w)}$ for which the method with still return each binary tree $T$ from any distance function on $X$ that differs by less than $\epsilon$ from $d$ on any pair of elements of $X$. In the case of Neighbor-Joining (and several other methods) this safety radius is $\epsilon= \frac{1}{2}w_{\rm min}$ where $w_{\rm min}$ is the smallest interior edge weight.  It is not hard to show that the safety radius of any tree reconstruction method based on distance data cannot be made any larger than this.

\subsection{Diversity measures.}

Given a phylogenetic $X$-tree $T$ with an edge weighting $w$, consider $L = \sum_ew(e)$, which is the total sum of the edge weights over the tree
Notice that for the tree in Fig.~\ref{tree5}(ii) we can write:
$$L= \frac{1}{2}[d(a,b)+d(b,c)+d(c,e)+d(e,f)+d(f,g)+d(g,h)+d(h,a)],$$
since the cyclic permutation $(abcefgh)$  traverses the tree in a clockwise order and so covers every edge exactly twice.
However,  there are other ways to embed the tree in the plane and do this -- for example, the cyclic
permutation $(aefcbhg)$ also traverses a different planar embedding of this tree in a clockwise order. 
It is easily shown that for any phylogenetic $X$-tree, the number $N_T$ of cyclic permutations 
that traverse the tree in a clockwise order is given by:
$$N_T=\prod_{v\in I(T)} ({\rm deg}(v)-1),$$
where ${\rm deg}(v)$ is the degree of vertex $v$, and $I(T)$ is the set of interior vertices of $T$ \cite{semp}.
For example, for the tree Fig.~\ref{tree5}(ii)), $N_T=  3 \times 3 \times 2=18$.
 
For each cyclic permutation $(x_1, x_2, \ldots, x_n)$, 
we have $L= \frac{1}{2}[d(x_1,x_2) +d(x_2,x_3) + \cdots + d(x_{n-1},x_n) + d(x_n, x_1)].$
If we average these over all the $N_T$ cyclic permutations that traverse $T$ in clockwise order, we obtain:
\begin{equation}
\label{lameq} L = \sum_{\{x,y\}} \lambda_T(x,y) d(x,y),
\end{equation}
where $\lambda_T(x,y)$ depends just on the number and degrees of the vertices in $T$ on the path $p(T; x,y)$ between $x$ and $y$.
When $T$ is a binary tree, $\lambda_T(x,y)$ is $(\frac{1}{2})^k$, where $k=k(x,y)$ is the number of interior vertices in $p(T;x,y)$.  
For instance, in the case of the quartet tree in Fig.~\ref{tree5}(i),  this gives:
$$L= \frac{1}{2}d(x,y) + \frac{1}{2}d(w,z) + \frac{1}{4}(d(x,w)+d(x,z)+d(y,w) + d(y,z)).$$
For a general phylogenetic tree (not necessarily binary) $\lambda_T(x,y) =1/ \prod_{v\in p(T; x,y)} ({\rm deg}(v)-1).$

The identity (\ref{lameq}) suggests a new way to build phylogenetic trees from distances, which is called `Balanced Minimum Evolution' (BME) \cite{pau}.
Given a arbitrary distance function  (not necessarily a tree metric) $\delta$ on $X$, this method
score each phylogenetic $X$-tree $T$ by the value:
$$L_\delta(T) =   \sum_{\{x,y\}} \lambda_T(x,y) \delta(x,y),$$
and searches for  the tree $T$ that has the smallest $L_\delta(T)$ score. If $\delta$ has a tree representation on some tree $T$, then this tree 
has the smallest $L_\delta$ score; moreover, like Neighbor Joining, the BME method comes with a `safety radius'  (allowing $\delta$ to be just sufficiently close to $d_{(T,w)}$) provided, as before, that $T$ is binary.  Mathematical results also show how BME can be viewed as a weighted least squares method \cite{des}. 

As well as considering the total diversity of the tree $L$, we can also consider how much diversity is spanned by different subsets of leaves.  This measure  is called {\em phylogenetic diversity} (PD),  and is important for biodiversity conservation \cite{pur}, and other applications.  Formally, given a phylogenetic $X$-tree and a positive edge weighting $w$, we can associate to each  subset $Y$ of $X$ a positive value, denoted $PD(Y)$, equal to  the sum of the weights of the edges of the minimal subtree of $T$ that connect the leaves in $Y$.  For example,  $L = PD(X)$, and $d_{(T,w)}(x,y) = PD(\{x,y\})$. And just as the PD scores of subsets of size $k=2$ (i.e. distances) can be used to reconstruct a tree, so can the PD scores of
subsets of size $k$ for any $k$ up to (but not exceeding) $\lceil n/2\rceil$ \cite{pac}.

The function $PD$  is clearly monotone -- the PD of a set is always greater than the PD of any strict subset; moreover PD  enjoys a `strong exchange' property: For any subset $Y_1$ of $X$ of size at least two, and any subset $Y_2$ of $X$ that is larger in size than $Y_1$ there always exists an element $y \in Y_2-Y_1$ for which:
$$PD(Y_1\cup \{y\}) + PD(Y_2 -\{y\}) \geq PD(Y_1)+PD(Y_2).$$
This property justifies a simple and fast strategy for finding a subset $Y$ of $X$ of any given size $k$ having maximal PD for sets of that size.  The strategy is simply the greedy one: first select two leaves $x,y$ that are furtherest apart in the tree (i.e. maximize $d_{(T,w)}(x,y)$) and then  sequentially add a leaf that increases the PD score by the maximum amount to the tree so-far constructed, until $k$ leaves are present.  Formally, the collection of subsets of $X$ that have maximal PD score for their cardinality form what is known in combinatorics as a `greedoid'.

A more sophisticated mathematical approach to the study of distances `T-theory' (tight-span), and split decomposition theory, pioneered by Andreas Dress and colleagues \cite{dre}, and 
an extension of this approach to diversity has recently been described \cite{bry}.

Distances and diversities also have a clear meaning if we weight the edges of a tree by arbitrary real values (possibly negative), or more generally by nonzero elements of an arbitrary Abelian group.  Several of the main results above extend with minor modification.  There is one `fly in the ointment'  however -- for distances, problems arise if the group has elements of order 2 (for instance, uniqueness of the tree representation fails, since all 15 phylogenetic trees having the shape shown in Fig.~\ref{shapesfig}(e) with edges assigned the element $1$ of $\cG=(\{0,1\}, +)$  induce exactly the same `distance' function).  But uniqueness  can be restored by moving to from distances to diversities, where not just pairs, but also triples of leaves are considered  \cite{dreste}.

\section{Markov models and the `Felsenstein Zone'.}
\label{markovsec}
 A major advance in phylogenetics has been the development of stochastic models to describe the evolution of genetic sequences  and genomes on a tree.  For genetic sequences, 
 these models typically describe point substitutions that occur at sites in the DNA sequence that  codes for some particular gene.  Such models allow biologists to convert the sequences we observe today at the leaves of the tree into an estimate of the tree itself  (and perhaps its branch lengths, or ancestral states within the tree).  By combining these `gene trees' one can in turn estimate the `species tree'.
 
 The rise of `statistical phylogenetics'  was pioneered in the 1960s and 70s  by Anthony Edwards, Joseph Felsenstein and others (including David Sankoff, with a visionary paper in this journal \cite{san}).  Today's methods of choice are based on maximum likelihood and Bayesian approaches.  Stochastic models assume  that characters evolve independently  on a tree, and the evolution of each character is described by some Markovian process; this may be the same across the characters or vary (for instance, some characters may evolve more rapidly than others).  

One of the catalysts that ushered in this stochastic approach was a landmark 1978 paper by Joseph Felsenstein \cite{fels2}.   He showed that if characters evolve independently  under a simple stochastic process then existing methods like maximum parsimony (discussed above) can be seriously misled. So, as the number of characters increases, it would  be increasingly certain that the maximum parsimony tree will be a different tree from the `true' tree (i.e. the one on which the characters evolved).  By contrast, other methods (like maximum likelihood) are, under certain conditions, provably statistically consistent and so converge on the true tree as the number of characters grows.

Felsenstein considered a simple process involving just two states -- let's call them $\alpha$ and $\beta$ --  which can flip between states with equal probability.  This process is familiar in coding theory as the `binary symmetric channel'.  In phylogenetics, we apply this this process to the edges of a tree -- each edge $e$ of the tree has a certain probability $p_e$ of a change of state between its endpoints, and, as in coding theory,  it is assumed that $p_e$ lies strictly between 0 and 0.5.   The model also assumes that the (marginal) state at any given leaf is uniform (i.e. no state is `preferred') and that changes of states on different edges are independent events.

\begin{figure}[htb]
\centering
\includegraphics[scale=1.2]{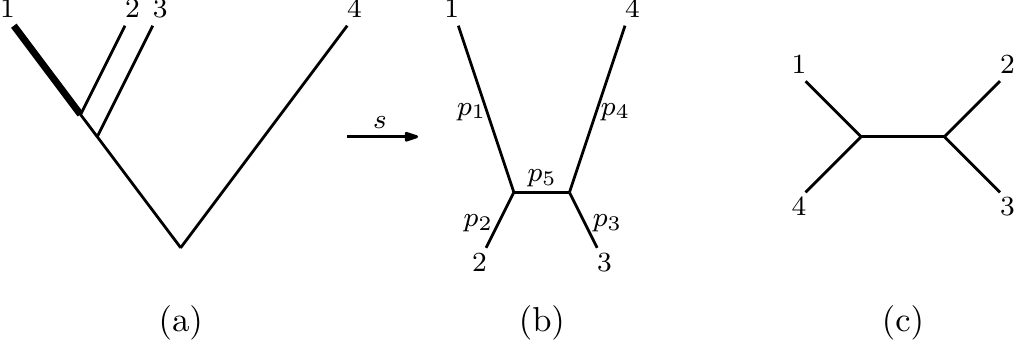}
\caption{(a) A high rate of evolution on the lineage leading to species $1$ and a distant outgroup species ($4$) can be modelled by a Markov process on the associated unrooted tree (obtained by suppressing the root) in (b); for this tree $T$,  if $p_1$ and $p_4$ are large enough relative to the other $p_i$ values, the maximum parsimony tree for a large number of characters generated on $T$ is likely to  be the tree $T'$ shown in (c).}
\label{fels}
\end{figure}

Felsenstein's tree is shown in Fig.~\ref{fels}(b) -- we can imagine it as a tree in which there has been an accelerated rate of evolution (resulting in higher probabilities of change) in two non-adjacent lineages. It can also be realized on a rooted tree as in Fig.~\ref{fels}(a), with a single rate increase in one short branch (the branch leading to $1$), and a distant out group species ($4$).   Denote the probabilities of change on the edges of the tree in Fig.~\ref{fels}(b) by the values $p_1, \ldots, p_5$, as shown.   

Now, there are $2^n=16$ different ways to assign the two states to a set $X$ of size $n$, but if we identify complementary assignments, obtained by interchanging $\alpha$ and $\beta$  (these two assignments have equal probability under the model) we get just  $2^{n-1}$ distinct {\em patterns}.  For a subset $A$ of $\{1,2,\ldots, n\}$, let $p_A$ be the probability of generating a pattern at the leaves of the tree in which $A$ is precisely  the leaves that are in different state to leaf $n$.  For example,  $p_\emptyset$ is the probability that all leaves are in the same state (i.e. all $\alpha$ or all $\beta$).   For the tree in Fig.~\ref{fels}(b), it is easily checked that:
$$p_\emptyset = (1-p_1)(1-p_2)(1-p_3)(1-p_4)(1-p_5) + p_1p_2p_3p_4(1-p_5)$$
$$+ p_1p_2p_5(1-p_3)(1-p_4) + p_3p_4p_5(1-p_1)(1-p_2). $$

There are various ways to compute the $p_A$ values, but one particularly elegant way that holds for any phylogenetic tree with $n$ leaves is by the following identity:
\begin{equation}
\label{hada}
p_A =\frac{1}{2^{n-1}}  \sum\limits_{\substack{B \subseteq \{1,2, \ldots n-1\} \\ |B|\equiv 0 \mbox{ mod 2}}} (-1)^{|A \cap B|}   \prod_{e \in P(T, B)} (1-2p_e)
\end{equation}
where $P(T,B)$ is the unique set of edges of $T$ that lie on any set of edge-disjoint paths in $T$ that connect pairs of leaves in the even cardinality set $B$.
For the tree in Fig.~\ref{fels}(b),  if we   let $x_i = (1-2p_i)$, and  take $A=\emptyset$ (so that $(-1)^{|A\cap B|}=1$ for all $B$ in 
Eqn.~(\ref{hada})) then we obtain:
\begin{equation}
p_\emptyset = \frac{1}{8}(1 + x_1x_2 + x_3x_4  +x_1x_3x_5+ x_2x_3x_5+ x_1x_4x_5+ x_2x_3x_5 + x_1x_2x_3x_4).
\label{emptyeq}
\end{equation}
All other $p_A$ values are obtained from the right-hand side of Eqn. (\ref{emptyeq}) by replacing $+$ by $-$ for exactly half the terms.
The somewhat mysterious representation in Eqn. (\ref{hada})  follows from a combinatorial study of this model (in which $ [(-1)^{|A \cap B|}]$ is a $2^{n-1} \times 2^{n-1}$ Hadamard matrix) due to Mike Hendy \cite{hen}, and later generalized  to other models using group representation based approaches  by Evans and Speed \cite{ev}, and Sz{\'e}kely {\em et al.} \cite{sze}. 

With this in hand, we can now establish the main ingredient in Felsenstein's classic result for maximum parsimony.

\begin{theorem}
\label{felthm}
For a character  generated on tree $T$ in Fig.~\ref{fels}(b)  under the 2-state symmetric model with $p_1=p_4 = P$, and $p_2=p_3=p_5 = Q$,  the expected parsimony score of $T$ is larger  than for the tree $T'$ in Fig.~\ref{fels}(c) precisely when $P^2 > Q(1-Q)$.
\end{theorem}

{\em Proof: }
The only 2-state characters that have different parsimony scores on $T$ and $T'$ are $f_{12}$ and $f_{23}$ where $f_{12}(1)=f_{12}(2)\neq f_{12}(3)=f_{12}(4)$ and
$f_{23}(2)=f_{23}(3) \neq f_{23}(1)=f_{23}(4)$.  Notice that $f_{12}$ has a parsimony score of 1 on $T$ and 2 on $T'$, while $f_{23}$ has a parsimony score of 1 on $T'$ and 2 on $T$.
Moreover, the probabilities of generating $f_{12}$ and $f_{23}$ under the 2-state symmetric model are $p_{12}$ and $p_{23}$ respectively (where $T$ is the generating tree).
Thus, if $\Delta$ denotes the parsimony score of a character (generated on $T$) on $T$ minus the parsimony score of that character on $T'$ then the expected value of $\Delta$, denoted $\EE[\Delta]$, satisfies:

\begin{equation}
\label{Edelta}
\EE[\Delta] = p_{23}-p_{12}.
\end{equation}
Now, if we let $x_i = 1-2p_i$, then  Eqn.~(\ref{hada}) for $n=4$, and $A=\{1,2\}$ and $\{2,3\}$ gives:
$$p_{12} = \frac{1}{8} (1 + x_1x_2 + x_3x_4 - x_1x_3x_5- x_2x_3x_5- x_1x_4x_5-x_2x_4x_5 + x_1x_2x_3x_4), \mbox{ and }$$
$$p_{23} = \frac{1}{8} (1 - x_1x_2 - x_3x_4  - x_1x_3x_5+ x_2x_3x_5+ x_1x_4x_5-x_2x_4x_5+x_1x_2x_3x_4).$$
Substituting these identities into Eqn.~\ref{Edelta} gives:
$$\EE[\Delta] = \frac{1}{4}(-x_1x_2-x_3x_4 + x_2x_3x_5+x_1x_4x_5).$$
Now, setting $x_1=x_4=u= (1-2P)$ and $x_2=x_3=x_5=v= (1-2Q)$ we obtain:
$$\EE[\Delta] = \frac{v}{4}[u^2+v^2-2u] = v(P^2-Q(1-Q)),$$
and so $\EE[\Delta]>0$ precisely if $P^2 > Q(1-Q).$
This completes the proof.
\hfill$\Box$

\bigskip

 Theorem ~\ref{felthm}, together with the Law of Large Numbers (or the Central Limit Theorem), ensures that for $k$ characters generated by the $T$ (with these $p_i$ values), a different tree, namely $T'$, will have a lower parsimony score than $T$, with probability converging to 1 as $k$ grows.  Intuitively, parallel changes on the two long branches of $T$ become more probable than a single change on the short edges. So, through the eyes of parsimony,  it is more optimal to join these two edges together in the reconstruction. This phenomenon of `long branch attraction'  has been observed in biological data \cite{hue}.

While parsimony can fail to recover the true tree, there are statistically consistent methods for inferring it.  A particularly simple one relies on the following distance function on $X$.  
For $x,y \in X$, let  $$\hat{\mu}(x,y) = -\frac{1}{2} \log(1 - 2\hat{p}(x,y)),$$ 
where $\hat{p}(x,y)$ is the proportion of characters that assign different states to $x$ and $y$.
Then provided we apply a distance-based tree reconstruction method with a positive safety radius ({\em c.f.} Section~\ref{met}) -- we are guaranteed to recover the underlying (unrooted) tree from $k$ independently evolved characters, as $k$ grows.  The reason is that,  as $k \rightarrow \infty$, the law of large numbers ensures that $\hat{p}(x,y)$ will converge to the probability $p(x,y)$  that leaves $x$ and $y$ are in different states, and so $\hat{\mu}(x,y)$ converges to $\mu(x,y)= -\frac{1}{2} \log(1 - 2p(x,y))$. It is then an nice exercise to show that  $\mu$ has a tree representation on the true tree $T$ with the edge weighting
$w(e) = - \frac{1}{2}\log(1-2p_e).$  That is, $\mu = d_{(T, w)}$.   The implication in (\ref{unique}) then ensures the reconstruction of both the unrooted tree  and the edge weights from $\mu$ (and thereby $\hat{\mu}$ for $k$ sufficiently large).

Biologists deal with much more complex models of character evolution than the  2-state symmetric model,  often on 4, 20 or 64 states (corresponding to DNA, amino acid and codon sequences, respectively).   For a general Markov model involving any state space, there is a way to construct a metric that has a tree representation on $T$, by taking the negative of the logarithm of the determinant of  the matrix of the joint probabilities of states for each pair of species.  In this way, the tree is identifiable from the probability distribution of characters.  This is enough to ensure that methods like maximum likelihood are statistically consistent.  However, for mixtures of such processes, the identifiability of the tree can easily be lost (mixtures of Markov processes are generally no longer Markovian).  This can be important for biologists -- if there are too many parameters to estimate from the data, then one may lose the ability to infer the one(s) we are interested in (such as the tree).   A striking example of this loss was provided for the 2-state symmetric model \cite{mat}:  if 50\% of DNA sites evolve on a 4-species tree $T$ with one carefully chosen set of branch lengths, and 50\% evolve on the same tree under a different chosen collection of branch lengths then the expected proportion of site patterns is exactly identical to that in which all sites evolve on a different tree with appropriately chosen edge lengths.

To obtain a deeper understanding of Markov processes on trees, techniques from commutative algebra and Lie algebra theory have proved invaluable  \cite{all, sum, sum2}. In particular, these techniques can be applied to determine the extent to which trees and other parameters of the model can be reconstructed from data (the `identifiability' issue mentioned above) \cite{all2}, a topic that is part of a broader emerging area called `algebraic statistics' \cite{drt}. The combinatorial topology and geometry of  two different notions of `tree space' are also of interest \cite{bil, mou}, as is the question of how much data we need to reconstruct a tree accurately.

\section{Current challenges.}
We have provided a brief overview of some of the central ideas in phylogenetics but much has been omitted and the reader interested in this area may wish to consult \cite{fel, sem} for further details. 

Two areas that are currently very active, and where mathematical and computational approaches play a key role include:

\begin{itemize}

\item  Using probability theory and combinatorics to study how the geneology of each gene (the `gene tree') for a set of species relates to the species'  phylogenetic tree (the `species tree'). Biologists typically now have very large numbers (thousands) of gene trees to compare species with, but these trees  can differ from the species tree by a process called `incomplete lineage sorting'. By considering how genes trace back in time and coalesce, it is possible to explain gene tree discordance and predict species trees from these conflicting gene trees (see e.g.  \cite{all2, deg, kno, mos}). 

\item
Extending phylogenetic tree theory to `phylogenetic networks' which are graphs that either display uncertainty in the data as to the likely species tree (implicit networks), or which provide an explicit representation of evolution where
there has been reticulation (such as the formation of hybrid species (see, for example, \cite{hus}). The patchy distribution of genes across taxa and lateral gene transfer also lead to further combinatorial and computational challenges \cite{roc}.

\end{itemize}

Finally, we have seen how any phylogenetic $X$-tree can be encoded by its associated set of splits,  and also by the leaf-to-leaf distances the tree induces under an edge weighting.  However, there is a third encoding, obtained by considering the quartet trees that are induced by the tree on subsets of $X$ of size four.   This association has led to some of the deepest results in phylogenetics (see e.g. \cite{gru}), and the exploration of the links between these three equivalent ways of encoding phylogenetic trees forms the basis of 
the emerging area of `phylogenetic combinatorics'  (for further details, see \cite{dre}).

\section{Acknowledgments.} 
Funding for this work was made possible by the NZ Marsden Fund and the Allan Wilson Centre.  I thank Simone Linz, Elliott Sober, Amelia Taylor and three anonymous reviewers for some helpful comments on an earlier version of this article. 

\bigskip

\noindent\textbf{Mike Steel} is  the director of the Biomathematics Research Centre at University of Canterbury, Christchurch New Zealand, where he teaches mathematics and statistics.  He is a  fellow of the Royal Society of New Zealand, and deputy director of the Allan Wilson Centre.   When he is not doing mathematics, he enjoys mountain running, biking and alpine climbing. 

\noindent\textit{School of Mathematics and Statistics, 
University of Canterbury, Christchurch, New Zealand\\
mike.steel@canterbury.ac.nz}

\end{document}